\documentclass[superscriptaddress,prl,letterpaper,10pt,onecolumn]{revtex4}
\usepackage{amsfonts}
\usepackage{amsmath}
\usepackage{amssymb}
\usepackage{graphics}

\input{tcilatex}
\begin{document}

\title{New bases for a general definition for the moving preferred basis.}
\author{Mario Castagnino}
\affiliation{CONICET, IAFE (CONICET-UBA), IFIR and FCEN (UBA), Casilla de Correos 67,
Sucursal 28, 1428, Buenos Aires, Argentina. Email:
castagninomario@gmail.com, phone: (5411) 4789-0179.}
\author{Sebastian Fortin}
\affiliation{CONICET, IAFE (CONICET-UBA) and FCEN (UBA), Casilla de Correos 67, Sucursal
28, 1428, Buenos Aires, Argentina. Email: sfortin@gmx.net, phone: (5411)
4789-0179.}
\pacs{03.65.Yz, 03.67.Bg, 03.67.Mn, 03.65.Db, 03.65.Ta, 03.65.Ud}

\begin{abstract}
\textbf{Abstract}: One of the challenges of the Environment-Induced
Decoherence (EID) approach is to provide a simple general definition of the
moving pointer basis or moving preferred basis. In this letter we prove that
the study of the poles that produce the decaying modes in non-unitary
evolution, could yield a general definition of the relaxation, the
decoherence times, and the moving preferred basis. These probably are the
most important concepts in the theory of decoherence, one of the most
relevant chapters of theoretical (and also practical) quantum mechanics. As
an example we solved the Omn\`{e}s (or Lee-Friedrich) model using our theory.
\end{abstract}

\keywords{Decoherence, preferred basis, relaxation tine, decoherence time.}
\maketitle

Decoherence is extremely important both for theoretical and applied physics.
In fact, decoherence is the main element of the quantum to classical limit
and this limit is one of the essential features of any successful
interpretation of quantum mechanics. But it is also essential for
technological subjects as for quantum computation. Nevertheless nowadays we
have not a generic definition of moving preferred basis, an essential
element for this theory.

In order to prove this assertion let us take into account the state of the
art on the subject: the book of Maximilian Schlosshauer \textquotedblleft
Decoherence and the quantum to classical transition\textquotedblright \cite%
{max}, where it is said that the fundamental equation $\langle
E_{i}(t)|E_{j}(t)\rangle \sim e^{-\frac{t}{\tau _{d}}}$ is valid for \textit{%
many environment models} (not for all), that the characteristic decoherence
timescale \textit{can be evaluated numerically} (i. e. not by a general
mathematical formula), and that \textit{frequently an exponential-decay
behavior }is found (not always). It is also said that a \textit{simple and
intuitive criterion will \ be introduced }to define the preferred set of
pointer states while a rigorous criterion would be necessary given the
importance of the subject. In fact the book presents a set of different
solved models but not a rigorous and \ general rule to define the moving
preferred basis. This also is the common lore according to the literature on
the subject.

Of course this is not a criticism to the immense scientific value of the
research already done on decoherence, but the justification in trying to do
a further step. And this step is also justified for both the theoretical and
practical importance on the subject and for the urgent need of a more
rigorous treatment.

This letter contains a tentative proposal for a general definition of the 
\textit{moving preferred basis} where the states of the (open) system become
diagonal in a certain decoherence time.

\textbf{a.-} Let us begin introducing the fundamental concepts:

\begin{itemize}
\item Relaxation is the (non-unitary) evolution of a quantum state $\rho (t)$
towards an equilibrium state $\rho _{\ast }$ at a relaxation time $t_{R}.$
At this time $\rho _{\ast }$ is obviously diagonal in its own eigen-basis $%
\{|i\ast \rangle \},$ the relaxation basis.

\item Decoherence is the (non-unitary) evolution of a quantum state $\rho
(t) $ towards a state where $\rho (t)$ is diagonal in some well defined
moving preferred basis $\{|i(t)\rangle \}$ at a certain decoherence time $%
t_{D}.$
\end{itemize}

The problem is that the last definitions are circular since these three
notions: decoherence, moving decoherence basis, and $t_{D}$ depend among
each other and none of them has a general independent definition.
Nevertheless there is a large number of convincing examples \cite{dec} where
these notions are defined, case by case. Often the moving decoherence basis
is related to some (macroscopic) collective variables which are
experimentally accessible \cite{OA}, and in this case we can call it a 
\textit{moving \textquotedblleft pointer\textquotedblright\ basis}. Using
these accessible variables an almost general definition can be found in \cite%
{OP}, but only in systems that satisfy a certain number of (reasonable)
conditions.

Obviously this state of affairs is not theoretically satisfactory so we need
a general definition of the three concepts: decoherence, moving preferred
basis, and $t_{D}$. In this letter all these notions are defined
independently and unambiguously \ but only for systems that satisfy the
usual properties of continuity and quasi-analiticity, i. e. the existence of
poles (but not more complex objects like branch-cuts, etc.). The poles
theory is widely used in quantum mechanics \cite{4A}, QFT \cite{4C}, and
especially in scattering theory \cite{4B}. Essentially, \ the states
associated with\textbf{\ }the poles or resonances are similar to plane
waves, both are non normalizable states. The plane waves would be the
eigenstates of the free Hamiltonian while the eigenstates corresponding to
the perturbed Hamiltonian would be the Gamov states \cite{Man} \cite{4D}.

\textbf{b.-} Thus we will try to introduce a new general definition that
would encompass the various definitions of \textquotedblleft
pointer\textquotedblright\ basis in a general one that we will simply call
the moving preferred basis. Based on this definition we will define
decoherence and decoherence time $t_{D}.$ The main idea will be the
following:

\begin{itemize}
\item Given a mechanical system we can find its typical oscillation modes.
Let us consider the Hamiltonian equations and find the constant of motion.
Then, the corresponding conjugated variables will evolve as $\varphi
_{i}=\varphi _{i}(0)+\omega _{i}t$ and these variables will define the
typical oscillation modes.

\item The quantum version of this mechanical example is obtained if we
quantize the system \cite{Ditto}. Then the typical modes would be related to
the complete set of commuting observables of the quantum Hamiltonian and
they would be $\exp (-\frac{i}{\hbar }\omega _{i}t)$ \ where $H|i\rangle
=\omega _{i}|i\rangle .$ Of course in this case the evolution of the system
is unitary.
\end{itemize}

To obtain non-unitary evolutions\ that would reach equilibrium and
decoherence we must consider some (relevant) observables $O_{R}=O_{S}\otimes
I_{E}$ where $O_{S}$ is any observable of the Hilbert space $\mathcal{H}_{S}$
of the (proper or relevant) system $\mathcal{S}$ and $I_{E}$ is the unit
operator of the Hilbert space $\mathcal{H}_{E}$ of the environment $\mathcal{%
E}$. Then if $(\rho |O)=Tr(\rho O)$ we have%
\begin{equation}
(\rho (t)|O_{R})=(\rho (t)|O_{S}\otimes I_{E})=(\rho _{S}(t)|O_{S})
\end{equation}%
where the last member is computed in the Hilbert space of system $\mathcal{S}
$ and $\rho _{S}(t)=Tr_{E}\rho (t),$ i. e. we have \textquotedblleft traced
away\textquotedblright\ the environment. $O_{S}\in \mathcal{O}_{S}$ is the
vector space of the observables of the system and\textbf{\ }$\rho _{S}(t)\in 
\mathcal{O}_{S}^{\prime }$,\ the dual space of\textbf{\ }$O_{S}$\textbf{.}
The $\rho _{S}(t)$ evolves non-unitarily.

Then to obtain the typical non-unitary modes we must extend the range of the
exponents of the evolution of the $(\rho _{S}(t)|O_{S})$ from the real
semiaxis to the complex plane obtaining the complex eigenvalues \cite%
{Complex}%
\begin{equation}
z_{i}=\omega _{i}-\frac{i}{2}\gamma _{i}
\end{equation}%
which also are the complex poles of the resolvent or those of the complex
extension of the S-matrix \cite{4A}. Then the characteristic decaying times
are $t_{i}=\frac{\hbar }{\gamma _{i}}$ and also a \textquotedblleft long
time\textquotedblright\ or \textit{Khalfin} decaying mode \cite{4A}. Usually
the corresponding time of this last mode is so long \cite{3A} that it can be
neglected for practical reasons, as we will always do below.

\textbf{c.-} Then it can be proved that the $\gamma _{i}$ define the
decaying modes \cite{Liotta}, since essentially%
\begin{equation*}
(\rho _{S}(t)|O_{S})=(\rho _{S\ast }|O_{S})+
\end{equation*}%
\begin{equation}
a_{0}\exp (-\frac{\gamma _{0}}{\hbar }t)+\sum_{i=1}^{N}a_{i}(t)\exp (-\frac{%
\gamma _{i}}{\hbar }t)  \label{-1}
\end{equation}%
where $\rho _{S\ast }$ is the equilibrium state of the proper system. Thus
it is quite obvious that the minimum of the $\gamma _{i}$, let us say $%
\gamma _{0},$ related to the pole $z_{0}$ closest to the real axis,
corresponds to the slowest decaying mode and therefore the \textit{%
relaxation time} is%
\begin{equation}
t_{R}=\frac{\hbar }{\gamma _{0}}  \label{0}
\end{equation}

Let us now go to the simplest example: a model with only two poles, $z_{0}$
and $z_{1}$, such that $\gamma _{0}\ll \gamma _{1}.$ Then we have%
\begin{eqnarray}
(\rho _{S}(t)|O_{S}) &=&(\rho _{S\ast }|O_{S})+a_{0}(t)\exp (-\frac{\gamma
_{0}}{\hbar }t)  \notag \\
&&+a_{1}(t)\exp (-\frac{\gamma _{1}}{\hbar }t)  \label{1}
\end{eqnarray}%
where\textbf{\ }$a_{0}(t)$ and\textbf{\ }$a_{1}(t)$ the are real oscillating
functions\textbf{\ }and, in this case, since we only have two modes,
necessarily $\gamma _{1}$ corresponds to the decoherence mode so%
\begin{equation}
t_{D}=\frac{\hbar }{\gamma _{1}}  \label{0'}
\end{equation}%
is a reasonable candidate for decoherence time. In fact, for $t>t_{D}$ the
third term of the l.h.s. of (\ref{1}) is negligible and we have%
\begin{equation}
(\rho _{S}(t)|O_{S})=(\rho _{S\ast }|O_{S})+a_{0}(t)\exp (-\frac{\gamma _{0}%
}{\hbar }t)  \label{1''}
\end{equation}%
Thus let us \ define a \textit{preferred state} $\rho _{P}(t),$ for all
times, such that%
\begin{equation}
(\rho _{P}(t)|O_{S})=(\rho _{S\ast }|O_{S})+a_{0}(t)\exp (-\frac{\gamma _{0}%
}{\hbar }t)  \label{1'}
\end{equation}%
According to the Riezs theorem \cite{Ballentine}\ the inner product $(\rho
_{P}(t)|O_{S})$\ defines the functional\textbf{\ }$(\rho _{P}(t)|\in
O_{S}^{\prime }$\textbf{. }$\rho _{P}(t)$ would be defined \textit{for all
times }and would be self adjoint since the rhs of the last equation is real.
Then we can find the eigen-decomposition of $\rho _{P}(t):$%
\begin{equation}
\rho _{P}(t)=\sum_{i=1}^{D_{S}}\rho _{i}(t)|i(t)\rangle \langle i(t)|
\label{2}
\end{equation}%
where\textbf{\ }$D_{S}$ is the dimension of the space $\mathcal{H}_{S}$.
Comparing eqs. (\ref{1}) to (\ref{1'}) it is quite evident that if we define 
$t_{D}=\frac{\hbar }{\gamma _{1}}$ , $\rho _{S}(t)$ becomes $\rho _{P}(t)$
for $t>t_{D}$.Then in this period $\rho _{S}(t)$ becomes diagonal in the
basis\textbf{\ }$\{|i(t)\rangle \}$ which is the only possible \textit{%
moving preferred basis} and $t_{D}=\frac{\hbar }{\gamma _{1}}$ is the 
\textit{decoherence time}.

\textbf{d.-} Going now to a the general case and therefore with an arbitrary
number $N$ of poles and therefore $N$ decaying modes $\gamma _{0}<\gamma
_{1}<\gamma _{2}...<\gamma _{N}$ and again we would be forced to say that
the relaxation time corresponds to the pole placed nearest to the real
semiaxis (then we obtain \ eq. (\ref{0})). \ Now we know that the main
achievement of the EID program is to obtain a decoherence time such that $%
t_{D}\ll t_{R}$ and as%
\begin{equation*}
\sum_{i=1}^{N}a_{i}\exp (-\frac{\gamma _{i}}{\hbar }t)=\exp g(t)=\exp
(g(0)+g^{\prime }(0)t+\frac{1}{2}g^{\prime \prime }(0)t^{2}+...
\end{equation*}%
and since we are only interested in small times we can only consider the two
first terms: i.e. $g(0)+g^{\prime }(0)t$ where%
\begin{equation*}
g(0)=\log \sum_{i=1}^{N}a_{i}(0),\text{ }g^{\prime }(0)=-\hbar ^{-1}\frac{%
\sum_{i=1}^{N}a_{i}(0)\gamma _{i}}{\sum_{i=1}^{N}a_{i}(0)}
\end{equation*}%
Then we have%
\begin{eqnarray}
(\rho _{S}(t)|O_{RS}) &=&(\rho _{S\ast }|O_{S})+a_{0}\exp (-\frac{\gamma _{0}%
}{\hbar }t)  \notag \\
&&+a_{eff}\exp (-\frac{\gamma _{eff}}{\hbar }t)
\end{eqnarray}%
and expression similar to eq. (\ref{1}) but where%
\begin{equation*}
\gamma _{eff}=\frac{\sum_{i=1}^{N}a_{i}(0)\gamma _{i}}{\sum_{i=1}^{N}a_{i}(0)%
}
\end{equation*}%
then in this general case we can define

\begin{equation}
t_{D}=\frac{\hbar }{\gamma _{eff}}
\end{equation}%
and%
\begin{equation*}
(\rho _{P}(t)|O_{S})=(\rho _{S\ast }|O_{S})+a_{0}(t)\exp (-\frac{\gamma _{0}%
}{\hbar }t)+
\end{equation*}

\begin{equation}
\sum_{i=1}^{M}a_{i}(t)\exp (-\frac{\gamma _{i}}{\hbar }t)
\end{equation}%
where in the last sum \ the term labelled by $i\leq M\leq N$ are those such
that $\gamma _{i}<\gamma _{eff}.$ Repeating the reasoning of point c in this
general case we can say that the moving preferred basis $\{|i(t)\rangle \}$
is the one that diagonalizes the new $\rho _{P}(t).$

This would be our candidate for the general definition of moving pointer
basis, that must be compared to those of the different models in the
literature \cite{dec}. The characteristic properties of this candidate
preferred basis are:

\begin{enumerate}
\item[i.] $\rho _{S}(t)$\textbf{\ }do decohere in this basis at time $t_{D}$%
\textbf{.}

\item[ii.] If we classify the decaying modes in slow modes (such that\textbf{%
\ }$\gamma _{i}<\gamma _{eff}$\textbf{) }and fast modes (such that\textbf{\ }%
$\gamma _{i}>\gamma _{eff}$\textbf{), }the evolution of our basis could be
considered as \textquotedblleft adiabatic\textquotedblright\ since it only
contains the slow modes.
\end{enumerate}

\textbf{e.-} Moreover, up to this point the reader may consider that all the
presented structure is completely alien to the usual literature on the
subject. It is not so, and at the end this letter we will use our technique
in an important example: the Omn\`{e}s (or Lee-Friedrich) model with
Hamiltonian \cite{OA}%
\begin{eqnarray}
H &=&\omega a^{\dagger }a+\int \omega _{k}b_{k}^{\dagger }b_{k}dk  \notag \\
&&\mathbf{+}\int (\lambda _{k}a^{\dagger }b_{k}+\lambda _{k}^{\ast
}ab_{k}^{\dagger })dk  \label{3}
\end{eqnarray}%
where $a^{\dagger }$\ ($b_{k}^{\dagger }$) is the \textquotedblleft
creation\textquotedblright\ operator of the system (environment) (an
interesting study of this model can be found in\textbf{\ }\cite{NA}). The
initial condition for the system corresponds to the sum of two Gaussian
functions and the environment initial condition is the vacuum state.
Moreover the Wigner transform of this initial condition must correspond to
two mountain-like density functions separated by a distance $L$ (which will
allow us to consider another important ingredient: \textit{macroscopicity}).
One of the properties of Hamiltonian (\ref{3}) is that if $|\nu \rangle $ is
an $\nu $-mode state it is easy to see that $H|\nu \rangle $ is also a $\nu $%
-mode state so the evolution of the system conserves the \textit{number of
modes sectors }(or number of \textquotedblleft particles\textquotedblright\
sectors). Thus, we can decompose the problem in number of modes sectors, and
we can study the one mode sector and find the corresponding pole $z_{0}$,
which is produced by the interaction of the Hamiltonian (\ref{3}). Then $%
z_{0}$, up to the second order in $\lambda ,$ is \cite{LCIB}:%
\begin{equation}
z_{0}=\omega +\int \frac{\lambda _{k}^{2}dk}{\omega -\omega _{k}+i0}
\end{equation}%
where 
\begin{equation}
z_{0}=\omega _{0}^{\prime }-\frac{i}{2}\gamma _{0}
\end{equation}%
and%
\begin{equation}
\gamma _{0}=\pi \int n(\omega ^{\prime })|\lambda _{\omega ^{\prime
}}|^{2}\delta (\omega -\omega ^{\prime })d\omega ^{\prime }
\end{equation}%
and where $dk=n(\omega ^{\prime })d\omega ^{\prime }.$ So we conclude that
in this sector $t_{R}=\frac{\hbar }{\gamma _{0}}$.

Now, going to the many sectors case (and always neglecting the Khalfin
term), it is easy to prove that the poles of the Omn\`{e}s system are%
\begin{equation}
z_{n}=nz_{0},\text{ \ \ \ \ }n=1,2,3,...  \label{ZN}
\end{equation}%
so $z_{0}$ is the pole closest to the real axis, for the whole system, and,
in fact, $t_{R}=\frac{\hbar }{\gamma _{0}}$ in this system.\ Then we have
our first coincidence since this result coincides with the one obtained by
Omn\`{e}s in page 288 of \cite{OA}, and a first evidence that the definition
of $t_{R}$ can be used.

Also using this result we can define an effective Hamiltonian, that produces
the non-unitary evolution of the system, and if we neglect the Khalfin term,
we obtain%
\begin{equation}
H_{eff}=z_{0}a_{0}^{\dagger }a_{0}=z_{0}N  \label{HEFF}
\end{equation}%
where $a_{0}^{\dagger }$ $(a_{0})$ is the creation (annihilation) operator
of the poles and $N$ is the number of poles operator. As we will see, the Omn%
\`{e}s model contains only the poles terms given by eq. (\ref{ZN}) but not
the Khalfin term (very slow motion term). This, completely justified
approximation, forces us to introduce the non Hermitian Hamiltonian of eq. (%
\ref{HEFF}), as it is usual in many cases. The presence of a non Hermitian
Hamiltonian is a good indicator since we are looking for a non unitary
evolution.\textbf{\ }Then $\left\{ |n\rangle \right\} $ is the common
eigenbasis of $H_{eff}$ and $N.$ Therefore if we define the amplitude $%
A(t)=\langle \psi |\varphi (t)\rangle ,$ and if $|\psi \rangle
=\sum_{n}a_{n}|n\rangle ,$ $|\varphi (t)\rangle =\sum_{n}b_{n}|n(t)\rangle $
we have%
\begin{equation}
A(t)=\sum_{n}b_{n}a_{n}^{\ast }e^{-i\frac{nz_{0}}{\hbar }t}  \label{4}
\end{equation}%
Now if we consider that the initial state of the system is $|\Phi (0)\rangle
=a|\alpha _{1}(0)\rangle +b|\alpha _{2}(0)\rangle $ and $\rho _{S}(0)=|\Phi
(0)\rangle \langle \Phi (0)\rangle |$ where $|\alpha _{1}(0)$ and $|\alpha
_{2}(0)\rangle $ are to two \textit{coherent states} placed at $\alpha
_{1}(0)=\frac{m\omega _{0}^{\prime }}{\sqrt{2m\hbar ^{2}\omega _{0}^{\prime }%
}}x_{1}(0)$ and $\alpha _{2}(0)=\frac{m\omega _{0}^{\prime }}{\sqrt{2m\hbar
^{2}\omega _{0}^{\prime }}}x_{1}(0),$ $p_{1}(0)=p_{2}(0)=0$ of phase space
the non diagonal part of $\rho _{S}(t)$ is%
\begin{eqnarray}
\rho _{S}^{(ND)}(t) &=&ab^{\ast }|\alpha _{1}(t)\rangle \langle \alpha
_{2}(t)|  \notag \\
&&+a^{\ast }b|\alpha _{2}(t)\rangle \langle \alpha _{1}(t)|
\end{eqnarray}%
Then using eq. (\ref{4}), the following inner products can be computed:%
\begin{equation}
\langle \alpha _{1}(0)|\alpha _{1}(t)\rangle =e^{-|\alpha
_{1}(0)|^{2}}e^{|\alpha _{1}(0)|^{2}e^{-i\frac{z_{0}}{\hbar }t}}
\end{equation}%
\begin{eqnarray}
\langle \alpha _{1}(0)|\alpha _{2}(t)\rangle &=&e^{-\frac{\left\vert \alpha
_{1}(0)\right\vert ^{2}+\left\vert \alpha _{2}(0)\right\vert ^{2}}{2}%
}e^{\alpha _{1}(0)\alpha _{2}(0)e^{-i\frac{z_{0}}{\hbar }t}}  \notag \\
&=&\langle \alpha _{2}(0)|\alpha _{1}(t)\rangle
\end{eqnarray}%
\begin{equation}
\langle \alpha _{2}(o)|\alpha _{2}(t)\rangle =e^{-\left\vert \alpha
_{2}(0)\right\vert ^{2}}e^{\left( \alpha _{2}(0)\right) ^{2}e^{-i\frac{z_{0}%
}{\hbar }t}}
\end{equation}%
So we can also compute the time evolution of $\rho _{S}^{(ND)}(t)$. Then,
with no lost of generality, we can choose 
\begin{equation}
\alpha _{1}(0)=0,\text{ \ \ }\alpha _{2}(0)\frac{m\omega _{0}^{\prime }}{%
\sqrt{2m\hbar ^{2}\omega _{0}^{\prime }}}L
\end{equation}%
where $L$ is the distance between the centers of the initial positions of
the two coherent states.

Then it turns out that%
\begin{eqnarray}
\rho _{S}^{(ND)}(t) &=&(ab^{\ast }|\alpha _{1}(0)\rangle \langle \alpha
_{2}(0)|+  \notag \\
&&+a^{\ast }b|\alpha _{2}(0)\rangle \langle \alpha _{1}(0)|)e^{-\frac{1}{2}%
L^{2}\left( 1-e^{-\frac{\gamma }{\hbar }t}\right) }  \label{A}
\end{eqnarray}%
Thus, using now the recipe $g(0)+g^{\prime }(0)t$ we obtain that the , $\rho
^{(ND)}(t)$ vanishes when $t\rightarrow \infty $ as%
\begin{equation}
\rho _{S}^{(ND)}(t)\sim \exp \left( -\frac{m\omega _{0}^{\prime }}{2\hbar
^{2}}\gamma _{0}L^{2}t\right)
\end{equation}%
(see \cite{Bonachon}, from details). Therefore, in this case,\ the moving
preferred basis is $\{|\alpha _{1}(t)\rangle ,|\alpha _{2}(t)\rangle \}$ for
large $L$ and from our previous definition of relaxation time $t_{R}=\frac{%
\hbar }{\gamma _{0}}$ we find that the decoherence time \ is%
\begin{equation}
t_{D}=\frac{2\hbar ^{2}}{m\omega }\frac{t_{R}}{L^{2}}
\end{equation}%
namely the result of Omn\`{e}s in page 291 \cite{OA}. This is our second
coincidence. So the Omn\`{e}s result can be obtained with our pole
technique. In this case macroscopicity appears for large $L.$ Moreover, it
can be proved\textbf{\ (}see \cite{Bonachon})\textbf{\ }that the Omn\`{e}s
moving preferred basis coincides with ours for large $L$.

The coincidence of our proposal with the Omn\`{e}s model leads us to
consider that our moving preferred basis (that diagonalizes $\rho _{P}$) is
a good candidate for the moving preferred basis.

\textbf{f.-} Of course we are aware that, to\ improve our proposal\textbf{s}%
, more examples must be added, as we will try to do elsewhere, because we
also believe that we have a good point of depart. In fact, our group is now
studying our proposal in other models as the Brownian motion and spin
systems. The Brownian motion models are so similar to the Omn\`{e}s one that
we believe that we will find a similar result. For the spin model, in order
to use the analytical continuation theory it is necessary to approximate the
quasi-continuous spectrum to \ a continuous one. We have already studied the
conditions for this approximation in \cite{Discreto}. Probably the
coincidences that we have found in the Omn\`{e}s model could be a general
feature of the decoherence phenomenon and would allow us to obtain complete
general definitions. Essentially because, being the pole catalogue the one
that contains \textit{all the possible decaying modes} of the non unitary
evolutions, relaxation and decoherence must be included in this catalogue,
since they are non-unitary evolutions.

In conclusion, we have given a quite general definition of a moving
preferred basis, decoherence time, and of the relaxation time. The Omn\`{e}s
formalism, of references \cite{OA} and \cite{OP}, contains the most general
definition of moving preferred basis of the literature on the subject. Our
basis have another conceptual frame: the catalogue of decaying modes in the
non-unitary evolution of a quantum system. Finally we hope that our result
would open a new way to obtain a general and rigorous formalism for one of
the most important chapters of quantum physics.

\textbf{Acknowledgments}

We are very grateful to Roberto Laura, Olimpia Lombardi, Roland Omn\`{e}s
and Maximilian Schlosshauer for many comments and criticisms. This research
was partially supported by grants of the University of Buenos Aires, the
CONICET and the FONCYT of Argentina.

\end{document}